\documentclass[journal,10pt]{IEEEtran}
\usepackage{amsmath,amsfonts}
\usepackage{algorithmic}
\usepackage{algorithm}
\usepackage{array}
\usepackage[caption=false,font=normalsize,labelfont=sf,textfont=sf]{subfig}
\usepackage{textcomp}
\usepackage{stfloats}
\usepackage{url}
\usepackage{verbatim}
\usepackage{graphicx}
\usepackage{cite}
\usepackage{amssymb}
\usepackage{booktabs}
\begin{document}

\title{Rate Compatible LDPC Neural Decoding Network: A Multi-Task Learning Approach}%A Multitask learning

\author{Yukun Cheng, Wei Chen,~\IEEEmembership{Senior Member,~IEEE,} Lun Li, Bo Ai,~\IEEEmembership{Fellow,~IEEE}
        % <-this % stops a space
\thanks{}%\emph{corresponding author: Wei Chen, Bo Ai}}
% <-this % stops a space
\thanks{Yukun Cheng, Wei Chen and Bo Ai are with the State Key Laboratory of Advanced Rail Autonomous Operation, Beijing Jiaotong University, Beijing, 100044, China (e-mail: ykcheng,weich,boai@bjtu.edu.cn)

Lun Li is with the State Key Laboratory of Mobile Network and Mobile Multimedia, Shenzhen 518057, China (e-mail: li.lun1@zte.com.cn).}

% <-this % stops a space
%\thanks{Manuscript received April 19, 2021; revised August 16, 2021.}
}

% The paper headers
%\markboth{Journal of \LaTeX\ Class Files,~Vol.~14, No.~8, August~2021}%
%{Shell \MakeLowercase{\textit{et al.}}: A Sample Article Using IEEEtran.cls for IEEE Journals}

\IEEEpubid{0000--0000/00\$00.00~\copyright~2022 IEEE}
% Remember, if you use this you must call \IEEEpubidadjcol in the second
% column for its text to clear the IEEEpubid mark.

\maketitle

\begin{abstract}%背景方法结果结论
Deep learning based decoding networks have shown significant improvement in decoding LDPC codes, but the neural decoders are limited by rate-matching operations such as puncturing or extending, thus needing to train multiple decoders with different code rates for a variety of channel conditions. In this correspondence, we propose a Multi-Task Learning based rate-compatible LDPC decoding network, which utilizes the structure of raptor-like LDPC codes and can deal with multiple code rates. In the proposed network, different portions of parameters are activated to deal with distinct code rates, which leads to parameter sharing among tasks. Numerical experiments demonstrate the effectiveness of the proposed method.
Training the specially designed network under multiple code rates makes the decoder compatible with multiple code rates without sacrificing frame error rate performance.
\end{abstract}

\begin{IEEEkeywords}
Rate-compatible LDPC codes, 5G, parameter-sharing, neural LDPC decoder.
\end{IEEEkeywords}

\section{Introduction}
\IEEEPARstart 
{L}OW-DENSITY parity-check (LDPC) codes were rediscovered in 1996\cite{mackay1997near} and are still a favored research direction in the field of error correction codes\cite{9606216, 9374097, 9336349}.%\cite{9374097}\cite{9336349}.
The fixed code rate of conventional capacity-approaching LDPC codes restricts their applications, especially when the system demands flexible code rates e.g. hybrid automatic repeat request (HARQ) mechanisms. In the 5G New Radio (NR), rate-compatible LDPC codes with a nested structure\cite{5174525, 6484236} are adopted as the coding scheme for data channels in the 5G enhanced mobile broadband (eMBB)scenario\cite{3gpp.38.212}. 

The convergence rate of the iterative message-passing algorithms such as the belief-propagation (BP) algorithm and min-sum (MS) algorithm gets slower due to the short cycles in the parity-check matrix of LDPC codes.
Modified algorithms (e.g. Normalized-BP, Offset-BP) that lower the confidence of the node information can migrate the effect of short cycles\cite{1001666}. However, these modified algorithms have some parameters for normalization or offset, which are usually determined empirically. It has been shown that the empirically set parameters fail to achieve optimum performance under different codewords and channel conditions.

%可以用一句话讲的工作比如谁做了什么干了什么
In recent years, in view of the fact that deep learning (DL) has shown remarkable improvement in various tasks, researchers combine the powerful learning capabilities of DL with traditional LDPC decoding algorithms to reduce computing complexity. In \cite{7852251}, Nachmani et al. introduce learnable weights into the BP decoding algorithm, effectively mitigating the negative impact of short cycles. However, a large number of nonlinear and multiplicative operations in the BP algorithm lead to the high computational complexity of the unrolled neural decoder.
MS-based decoding networks (e.g NNMS and NOMS) are then proposed to further reduce the computational complexity and improve the hardware adaptability\cite{8006751, 9875067}. To deal with long LDPC code and the gradient vanishing issue, a parameter-sharing mechanism is introduced in \cite{9427170} to reduce the number of parameters of the neural decoder for PB-LDPC.

\IEEEpubidadjcol 

DL-based decoding methods assume some fixed code rate. In the 5G NR, flexible LDPC code rates are supported by exploiting rate-matching operations such as puncturing and parity-check matrix extension\cite{AHMADI2019411}. Although one could train multiple networks suited to a range of code rates that would then be selected during the decoding process, this would require considerable storage, which restricts its applicability in resource-constrained applications such as the internet of things (IoT). Therefore, it is desired to develop new neural decoding networks capable of adapting to varying LDPC code rates. 

In this correspondence, we design a new rate-compatible (RC) decoding network based on Multi-Task Learning (MTL). Utilizing the extension structure, the proposed RC decoder can adaptively adjust the network structure according to the code rate. The parameter-sharing mechanism allows one set of parameters to cope with multiple code rates thus effectively reducing storage cost and training complexity. %Notably, this method is suitable for building both BP decoders and MS decoders.

\section{Preliminaries}
\subsection{Conventional BP/MS for LDPC Decoding}
According to the parity-check matrix $H$, the Tanner graph defines two kinds of nodes: variable nodes (VNs) and check nodes (CNs). Each VN corresponds to a bit in codeword (a column of $H$) and, each CN corresponds to a parity-check equation (a row of $H$). An edge connects a VN and a CN corresponds to a ``1" in $H$.

The conventional BP algorithm and MS algorithm use log-likelihood ratio (LLR) as messages passing through the edges. Let $y$ and $x$ denote a received noisy codeword and its corresponding pure codeword, respectively.

%需要给公式加上标点符号
The LLR value of the $v$-th bit $x_v$ is defined as
\begin{equation}
\label{1}
L_v = \ln \frac{P(y_v|x_v=0)}{P(y_v|x_v=1)},
\end{equation}
where the $P(y|x)$ denotes the transition probability from $x$ to $y$. Let $e=(v,c)$ represent the edge connecting the $v$-th VN and the $c$-th CN. For the $l$-th iteration, the CN to VN (C2V) message passing on the edge $e$ and the VN to CN (V2C) message passing on the edge $e$ are defined as $C_e^l$ and $V_e^l$, respectively. 
The V2C messages $V_{e}^{l}$ is calculated using
\begin{equation}
\label{BPVC}
V_{e}^{l} = L_v + \sum_{e'\in\varepsilon_v \backslash c}C_{e'}^{l},
\end{equation}
where $\varepsilon_v$ denotes the set of edges which are adjacent to the $v$-th VN, and $\varepsilon_v \backslash c$ denotes the edge set excluding the edge adjacent to the $c$-th CN.
The C2V message $C_e^l$ is calculated using
\begin{equation}
\label{BPCV}
C_e^l = 2 tanh^{-1}\left(\prod_{e'\in\varepsilon_c \backslash v}tanh\left(\frac{V_{e'}^{l-1}}{2}\right)  \right),
\end{equation}
where $\varepsilon_c$ denotes the set of edges which are adjacent to the $c$-th CN, and $\varepsilon_c \backslash v$ denotes the edge set excluding the edge adjacent to the $v$-th VN. At the first iteration, $C_e^0$ is initialized to 0 in Eq. (\ref{BPVC}).
For the $l$-th iterations, the output is the LLR value of $x_v$, which is defined as
\begin{equation}
\label{BP_OUT}
o_v = L_v + \sum_{e\in \varepsilon_v}C_{e}^{l},
\end{equation}
then the $v$-th decoded bit $\hat{x}_v$ can then be obtained using 
\begin{equation}
\label{hard_decision}
\hat{x}_v = \frac{1-sgn(o_v)}{2}.
\end{equation}

It is worth noting that in the MS algorithm, Eq. (\ref{BPCV}) will be rewritten as
\begin{equation}
\label{MSCV}
C_e^l =\prod_{e'\in\varepsilon_c \backslash v}sgn\left(V_{e'}^{l-1}\right)\times\min_{e'\in \varepsilon_c \backslash v}\left|V_{e'}^{l-1}\right|.
\end{equation}

In the decoding process, the V2C message is calculated first, then the C2V information is calculated. When the break conditions are met, the algorithm stops iterative calculation and outputs the final decoding result.

\subsection{Rate-Compatible (RC) Raptor-Like LDPC Code} 
In the 5G NR, raptor-like LDPC (RL-LDPC) codes are employed.
The parity-check matrix of a RL-LDPC code can be decomposed to a precode submatrix and multiple incremental redundancy code (IRC) submatrices \cite{7045568}. The precode submatrix defines a low-threshold high-rate precode, and a lower-rate parity check matrix is obtained by extending the precode submatrix using IRC submatrices, which means adding the same number of columns (VNs) and rows (CNs) to the precode matrix. 

The connections between CNs in the IRC submatrix and VNs, including a degree-1 VN and several precoded VNs, allow efficient encoding of the incremental redundancy. The adaptive rate of RL-LDPC codes is achieved by performing a varying number exclusive-or operations on the precoded symbols to generate different numbers of parity-check bits, and high-rate codewords are nested in the low-rate codewords.

\subsection{Multi-Task Learning}
MTL trains a model that fits multiple related tasks. The model exhibits better generalization and performance on each task, and reduces the risk of overfitting under a training set including data of different tasks. 
To reduce storage costs and improve prediction accuracy, sharing of partial network structures as well as weights is applied in MTL applications\cite{9392366}.

%单栏图片，换大箭头，图片大小。boundingbox
%\begin{figure}[!t]
%\centering
%%\setlength{\abovecaptionskip}{-1cm}
%\centerline{\includegraphics[width=0.23\textwidth]{RC-LDPC.eps}}
%\caption{Parity-check matrix of RL-LDPC.}
%\label{fig_1}
%\vspace{-0.5cm}
%\end{figure}

\section{Proposed Rate Compatible Decoding Network}
In this section, we present details of the proposed RC decoding network. 
The existing LDPC neural decoders introduced in the previous section are limited to a predefined code rate, while the BP and MS decoding methods are not restricted to the code rate. Here, we propose a rate-compatible LDPC neural decoder using a special parameter-sharing mechanism. We utilize the structure of the raptor-like LDPC codes and construct an adjustable dynamic neural decoding network.
\subsection{Rate Compatible Model-Driven Decoding Network}

\subsubsection{Sturcture of the neural decoder}
\begin{figure}[!t]
\centering
\setlength{\abovecaptionskip}{-0.2cm}
\centerline{\includegraphics[width=0.5\textwidth]{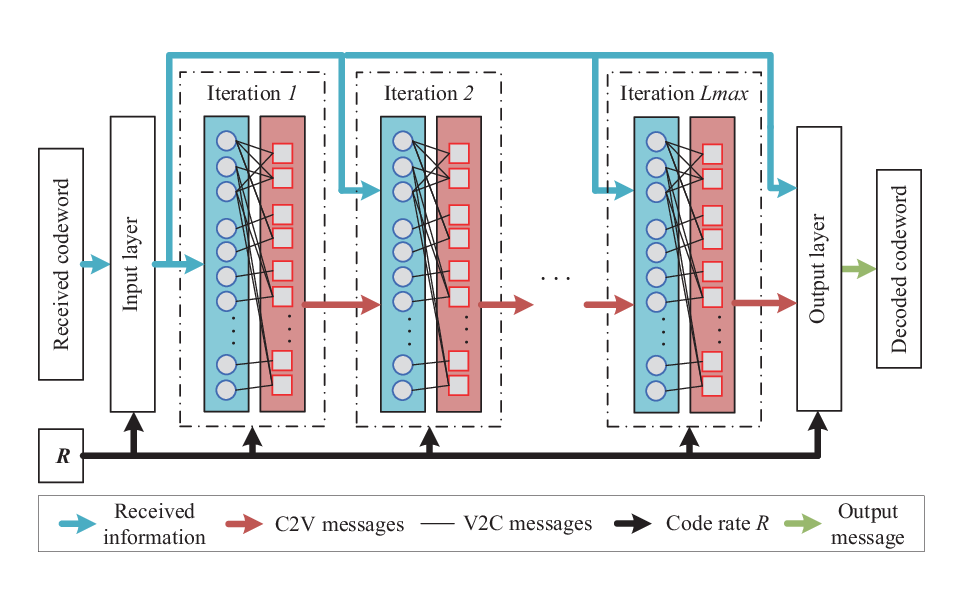}}
\caption{Sturcture of proposed decoding network.}
\label{fig_2}
\vspace{-0.5cm}
\end{figure}

The structure of our decoding network is shown in Fig \ref{fig_2}, which is constructed by unrolling the iterative decoding algorithm into a non-fully connected neural network. Each iteration in the conventional decoding algorithm is unrolled into a VN-sublayer and a CN-sublayer. Since the messages are propagated along the edges, each edge becomes a neuron in the sublayers.
The input layer is of size $N$, corresponding to the code length. Each hidden layer is of size $E$ (including all VN-sublayers and CN-sublayers), where $E$ denotes the number of edges in the Tanner graph. The size of the output layer is also $N$. To support different code rates, all layers adjust the number of neurons activated according to the code rate of the received code word, and thus the sharing of neurons for different tasks, i.e., code rates, leads to advantages brought by the MTL.

%In specific, similar to the message flow in the Tanner graph, each neuron outputs a message to the connected VN/CN node. For the $l$-th iteration, the VN-sublayer and CN-sublayer are represented as $l_{vn}$ and $l_{cn}$ respectively. For the first VN-sublayer, the neurons corresponding to the edges connecting the $v$-th variable node are associated with the $v$-th element in the input layer. For the layer $l_{vn}$, the neuron associated with the edge $e=(v,c)$ is connected to the neurons corresponding to the edge set $\varepsilon_c$. For the layer $l_{cn}$, the neuron corresponding to edge $e=(v,c)$ is connected with the neurons corresponding to the edge set $\varepsilon_v$.

\subsubsection{Message flow}
The message flow of the decoding network is shown as follows. 
For a received noisy codeword $y$, the input layer computes its LLR value $L_v$ and propagates this value to each VN-sublayer and output layer. For the $l$-th iteration, the V2C messages $V_e^l$ are calculated using Eq. (\ref{BPVC}),
where $C_e^0$ are all set to be 0 at the initialization step. 

%Short cycles in LDPC parity check matrix introduce autocorrelations to V2C messages (ideally uncorrelated), the computed C2V messages based on these messages are considered unreliable. To mitigate this issue, we utilize learnable parameters to adjust the confidence of the calculated C2V messages in Eq. (\ref{BPCV}), thereby migrating the defects brought by short cycles \cite{1001666}.
We add weights $w_e^l$ and biases $b_e^l$ to the C2V message passing on the edge $e$ in the $l$-th CN-layer, which leads to:
%Instead of using the C2V message in Eq. (\ref{BPCV}), we introduce learnable weights $w_e^l$ and biases $b_e^l$ to the C2V message passing on the edge $e$ in the $l$-th CN-layer, which leads to:
\begin{equation}%
\label{NNBPCV}
C_e^l = w_e^l\times2tanh^{-1}\left(\prod_{e'\in\varepsilon_c \backslash v}tanh\left(\frac{V_{e'}^{l-1}}{2}\right) \right)+b_e^l.
\end{equation}
%需要说一下
If setting weights and biases to 1 and 0, respectively, the decoding network is equivalent to the conventional BP decoder. 

The output of the network is the channel information corrected by the refined C2V messages, the equation is
\begin{equation}
\label{NNBPOut}
o_v = \sigma(L_v + \sum_{e\in \varepsilon_v}C_e^{L_{max}-1}),
\end{equation}
where $\sigma$ denotes the sigmoid activation function that maps the output into the range of [0,1]. 
With a hard decision on the output, we can get the predicted codeword $\hat{x}$.

It should be noted that if one unfolds the neural network based on the MS algorithm rather than the BP algorithm in the same way, then Eq. (\ref{NNBPCV}) should be rewritten as
\begin{equation}
\begin{split}
\label{NNMSCV}
C_e^l= & \prod_{e'\in\varepsilon_c \backslash v}sgn\left(V_{e'}^{l-1}\right) \\& \times {ReLU}\left(\min_{e'\in\varepsilon_c \backslash v}\!w_e^l\times \left|V_{e'}^{l-1}\right|+b_e^l\right),
\end{split}
\end{equation}
where $ReLU$ denotes the popular ReLU activation function. This design was proposed in \cite{9427170}. There are other neural LDPC decoding designs \cite{9145237, 8242643, 8006751} that employ different check node and variable node update equations. Note that the proposed rate-compatible mechanism is not coupled with the neural LDPC update designs, and can be applied to these different neural LDPC decoders. 

\subsubsection{Rate compatible parameter sharing mechanism}
\begin{figure}[t]
\centering
\setlength{\abovecaptionskip}{-0.5cm}
\centerline{\includegraphics[width=0.45\textwidth]{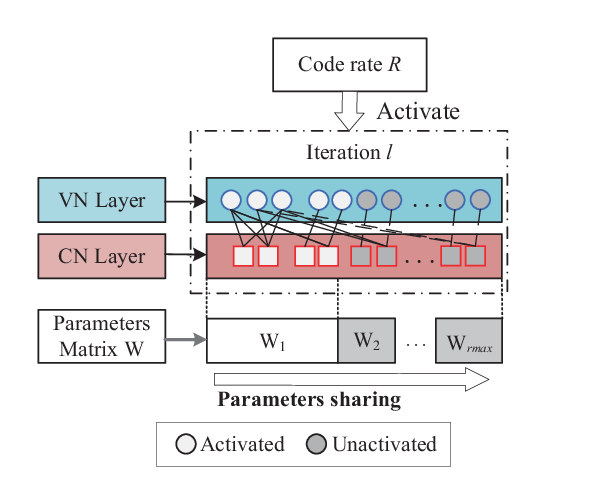}}
\caption{Parameter sharing mechanism in the proposed LDPC decoding network.}
\label{fig_3}
\vspace{-0.6cm}
\end{figure}
Since the network needs to support multiple different code rates, the hidden layer needs to adjust the width according to different code rates. Thus we propose a parameter-sharing mechanism, as shown in Fig. \ref{fig_3}, where the code rate controls the activation status of neurons in each sublayer. Owing to the raptor-like structure, the parameters of a high code rate are reused in the decoding process under a low code rate. 

The parameter matrix $W \in R^{2L_{max}\times E}$ of the decoding network is a hierarchical cascade of the weights and biases of each layer, where $R$ denotes the set of real numbers. The $(2l\!-\!1)$-th (even) rows and the $2l$-th (odd) rows of the parameter matrix, where $l=1,2,..., L_{max}$, serves as the weights and biases vector of the $l$-th layer, respectively.
As high-rate codewords are nested in the low-rate codewords, the whole network including all parameters is used for decoding codewords with the lowest rate, while only part of the network is activated for decoding codewords of the higher rate. The edge set $\varepsilon_c$ and $\varepsilon_v$ used in Eq. \eqref{BPVC},\eqref{NNBPCV}-\eqref{NNMSCV} are rewritten as $\varepsilon_c^R$ and $\varepsilon_v^R$, denoting the set of activated edges that are adjacent to the $c$-th CN and $v$-th VN under the code rate $R$, respectively.
This is the key of the proposed parameter-sharing mechanism.

Assuming the decoding network supports $r_{max}$ different code rates, the parameter matrix $W$ can be decomposed into $r_{max}$ submatrices, i.e., $W=[W_{1}, W_{2}, \ldots, W_{rmax}]$,
where the parameter matrix $W_1$ is used for decoding codewords of the highest code rate, $W_{1}$ and $W_{2}$ are used for decoding the codewords of the 2ed highest code rate, and ${W}_{i=1,...,t}$ are used for decoding the codewords of the $t$-th highest code rate.

We consider the decoding tasks at different code rates as related tasks and employ MTL to train a decoding model that can adapt to multiple rates. Specifically, by employing the proposed parameter-sharing mechanism, same parameters can cope with multiple code rates, which leads to the reduction of storage cost and training complexity compared to training multiple networks for different code rates.

The proposed scheme can be also seen as the result of another strategy for multiple rates, where the incoming LLRs are loaded into an LDPC coder with the lowest rate, with the punctured bits filled as zero LLRs. In this way, the punctured degree-one VNs only connect to a single CN. During the iterative decoding, the V2C messages of these VNs are generated solely from the received channel information. When filled with zero LLR, the C2V messages from the connected CN to other VNs all can be computed into zeros. In our proposed scheme, the corresponding neural nodes are inactivate, which is equivalent to C2V messages with zero LLRs.
\subsection{Training Method}
We use binary cross-entropy between the predict codeword and the transmitted codeword as the loss function, which is written as
\begin{equation}
Loss(\hat{x},x)=-\frac{1}{N}\sum_{v=1}^{N}x_vlog(\hat{x}_v)+(1-x_v)log(1-\hat{x}_v),
\end{equation}
where $\hat{x}_v$ and $x_v$ represent the $v$-th bit of the decoded message and the transmitted codeword, respectively. 
To learn parameters in a deep network, we adopt the layer-by-layer greedy training method.

In the MTL framework, the parameters involved in multiple decoding tasks are trained on various datasets corresponding to different code rates. Specifically, parameters used for decoding codewords of some code rate would be learned from all data with code rates no higher than that rate, e.g. the optimize step of $W_{i}$ can be written as $W_{i} = W_{i}-\sum_{t=i}^{r_{max}} \frac{\partial Loss_t}{\partial W_{i}}$
where $Loss_t$ denotes the decoding loss of the $t$-th highest code rate.
%Besides, the training set employs mixed codewords containing different SNR conditions.
 %the training SNR ranges from 0dB to 5dB with an interval of 0.5 dB. 

\section{Performance evaluation}%顺序需改改变，先讲场景再讲具体的
In this section, we construct neural RC decoders for the 5G NR LDPC codes and evaluate the frame error rate (FER) performance under the BPSK/QPSK modulation and the additive white Gaussian noise (AWGN) channel at different signal-to-noise ratios (SNRs). The experiments are conducted on the TensorFlow platform and GTX1050 GPU. Note that the proposed decoders do not have to run in GPUs, although the use of GPU significantly reduces the training time.

We employ the base code BG1 defined in 5G standard\cite{3gpp.38.212} with lifting factor $Z$=16 for BP-based decoders and MS-based decoders. The training set employs mixed codewords that are randomly generated under multiple SNRs ranging from 0dB to 6dB. Additionally, these codewords are generated with various code rates randomly selected from [0.2451, 0.3701, 0.5137], and these code rates are chosen from the 5G NR modulation and coding scheme (MCS) table for physical downlink shared channel (PDSCH)  \cite{3gpp.38.214}. 
%In our experiments, we filled the punctured bits with zero LLRs to make the codewords align with multiples of Z before inputting them into the network. Based on our previous analysis, these zero-filled LLRs do not affect the error performance of the decoding network.

The neural RC decoders have 20 hidden layers, which correspond to 20 full iterations of the BP/MS algorithm. For each update, the parameters are trained over 10000 batches, with 300 samples in each batch.
 We applied adaptive moment estimation (Adam) optimizer with an initial learning rate equal to 0.0001. To stabilize the operation of the learned BP decoder, the absolute value of the messages transferred between the neurons is clipped to 20\cite{7852251}. 

\subsection{Simulation Results}
To evaluate the performance of the proposed RC neural normalized BP (RC-NNBP) decoder and RC neural normalized MS (RC-NNMS) decoder, we conduct comparisons against the conventional iterative decoding algorithms (with 20 iterations) and the neural decoding networks trained under each code rate (with 20 hidden layers). For the BP-based decoders, the standard BP algorithm and neural normalized BP (NNBP) decoders are selected as competitors. For the MS-based decoders, the conventional normalized MS (CNMS) algorithm with weights 0.8 \cite{1495850}, and the neural normalized MS (NNMS) decoders are considered in the comparison. Note that early termination can be implemented in these neural decoders by examining the decoded bits after each iteration. For a fair comparison, we performed 20 iterations for all decoders.

\subsubsection{BP-based Decoders}

\begin{figure}[!t]
	\centering
	\setlength{\abovecaptionskip}{-0.3cm}
	\centerline{\includegraphics[width=0.35\textwidth]{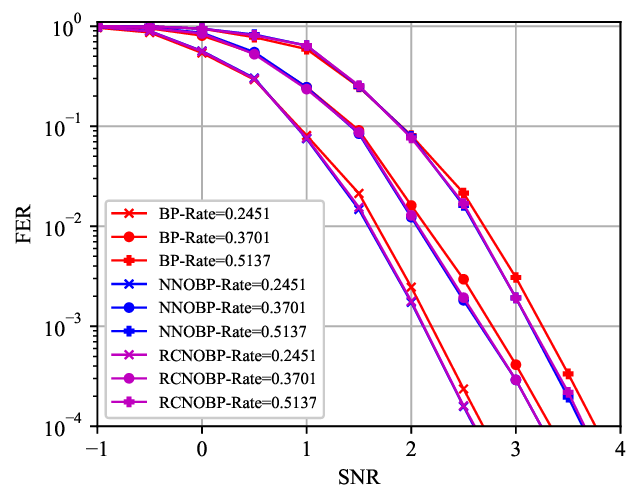}}
	\caption{Performance of RC-BP-Based decoder (Z=16).}
	\label{fig_4}
	\vspace{-0.5cm}
\end{figure}

The FER results of (160, 653), (160, 430), and (160, 311) BG2 codes are given in Fig. \ref{fig_4}, whose code rates are 0.2451, 0.3701 and 0.5137, respectively.
We observe that neural BP-based decoders outperform traditional BP algorithm. 
%At the code rate of 1/2, the neural decoder outperforms the BP algorithm by 0.3 dB. 
The proposed RC-NNBP decoder exhibits similar FER performance to NNBP decoders trained at different code rates, which in practice require multiple networks and result in greater storage costs.
\subsubsection{MS-based Decoders}

\begin{figure}[!t]
\centering
\setlength{\abovecaptionskip}{-0.3cm}
\centerline{\includegraphics[width=0.35\textwidth]{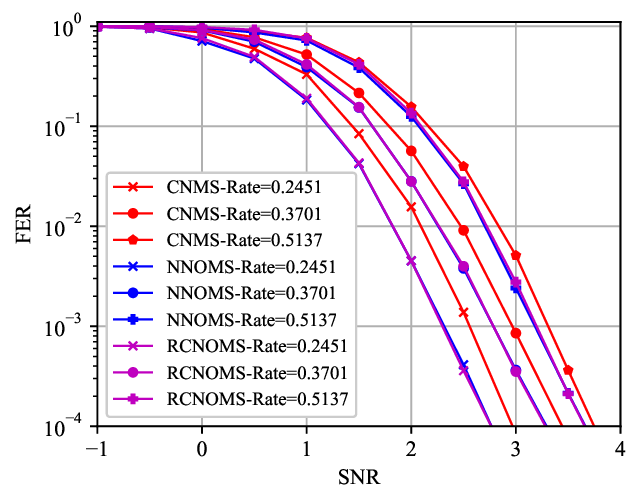}}
\caption{Performance of RC-MS-Based decoder (Z=16).}
\label{fig_5}
\vspace{-0.5cm}
\end{figure}

The FER results of MS-based decoders are given in Fig. \ref{fig_5}. We observe that the neural MS-based decoders exhibit superior performance compared to traditional algorithms, at the code rate of 0.2451, the neural decoder outperforms the CNMS algorithm by around 0.3dB. 
The proposed RC-NNOMS decoder also exhibits similar FER performance to NNOMS decoders,  which further demonstrates that the proposed RC decoder can effectively replace multiple conventional neural decoders, thereby reducing storage costs.
Understandably, a similar phenomenon can be seen with the BG1 codes.

\begin{figure}[!t]
	\centering
	\setlength{\abovecaptionskip}{-0.3cm}
	\centerline{\includegraphics[width=0.35\textwidth]{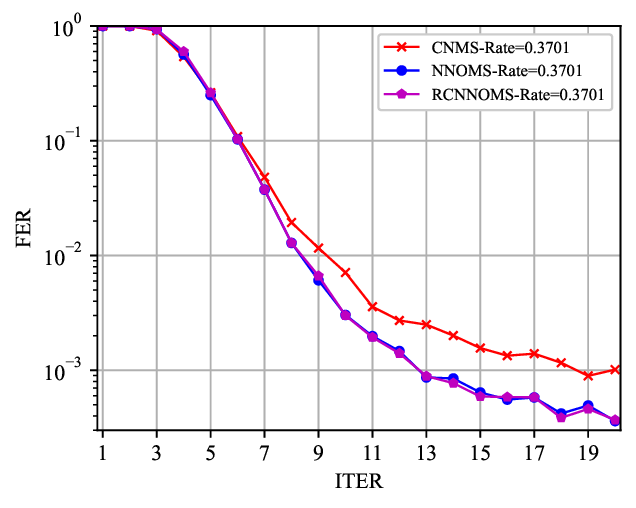}}
	\caption{FER-ITER performance of MS-Based decoders (Z=16, SNR=3dB).}
	\label{fig_6}
	\vspace{-0.5cm}
\end{figure}

Furthermore, the performance of the three MS-based decoders in terms of FER-iterations is presented in Fig. \ref{fig_6}. The experiment was conducted using an SNR of 3dB and the (160, 430) BG2 code. Remarkably, the neural decoder demonstrates superior convergence compared to the conventional algorithm. When aiming for an FER of $10^{-3}$, the neural decoder requires 6 fewer iterations to achieve it.

\subsubsection{Protograph-Based Rate Compatible (PBRC) Decoder}
The protograph-based LDPC (PB-LDPC) codes are adopted in 5G NR. By exploiting the lifting structure of PB-LDPC codes, \cite{9427170} proposed a method to deal with different lifting factors with one neural decoder, which uses the same parameter among a bundle of edges belonging to the same edge in the base graph. This approach in \cite{9427170} can be integrated into our method to deal with possible lifting factors and further reduce storage cost and training complexity. 

\begin{figure}[!t]
\centering
\setlength{\abovecaptionskip}{-0.3cm}
\centerline{\includegraphics[width=0.35\textwidth]{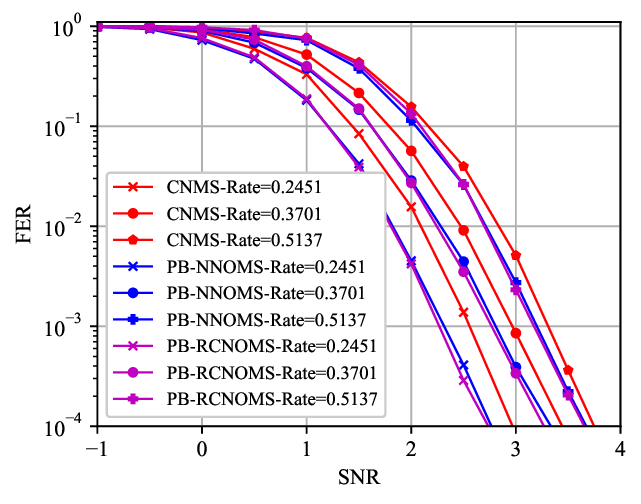}}
\caption{Performance of PBRC-MS-Based decoder (Z=16).}
\label{fig_7}
\vspace{-0.4cm}
\end{figure}

Figs. \ref{fig_7} show the FER results of the PBRC Decoders, where the code configurations are the same as that in Fig. \ref{fig_5}. The results indicate that our rate-compatible mechanism still proving useful in the PB-LDPC decoders.

The high performance of our proposed neural decoder benefits from two aspects: the well-learned parameters mitigating the effects of short cycles in the parity-check matrix, which leads the decoder to a faster convergence; the other is that the rate-compatible mechanism accommodates the decoder to multiple code rates, thus achieve a better FER performance and a lower storage cost at the same time.

\subsection{Complexity}

\begin{table}
	\renewcommand{\arraystretch}{1.5}
	\begin{center}
		\label{tab1}
		\caption{Operation and storage costs of each iteration}
		\resizebox{0.48\textwidth}{0.8in}{
		\begin{tabular}{| c | c | c | c | c | c | c |}
			\hline
			Operation & tanh & $\times$ & + & comp. &sign flip & Storage\\
			\hline
			BP& $2E$ & $\approx 2E$ & $\approx 2E$ & - & - & - \\
			\hline
			CNMS& - & $E$ & $\approx 2E$ & $\approx 2E$ & $\approx 2E$ & 1 \\ 
			\hline
			NNOBP& $2E$ & $\approx 3E$ & $\approx 3E$ & - & - & $\sum\limits_{i=1}^3{E_i}$ \\
			\hline
			RC-NNOBP& $2E$ & $\approx 3E$ & $\approx 3E$ & - & - &$E_1$  \\
			\hline
			NNOMS& - & $E$ & $\approx 3E$ & $\approx 2E$ & $\approx 2E$ & $\sum\limits_{i=1}^3{E_i}$ \\ 
			\hline 
			RC-NNOMS& - & $E$ & $\approx 3E$ & $\approx 2E$ & $\approx 2E$ & $E_1$ \\ 
			\hline 
		\end{tabular}
	}
	\end{center}
	\vspace{-0.5cm}
\end{table}

The required operation and storage costs in one iteration of different schemes are presented in Table \ref{tab1}, where $E_1, E_2$ and $E_3$ represent the total activated numbers of the neurons under code rate 0.2451, 0.3701 and 0.5137, respectively. 
The neural decoders have similar processing procedure in each iteration as BP and CNMS. 
The proposed RC decoders, i.e., RC-NNOBP/RC-NNOMS, retain the same computing complexity as conventional neural decoders, i.e., NNOBP/NNOMS, but require less memory to store the parameters to deal with multiple code rates.
According to the hardware implementation in \cite{9643510}, the weights and biases in the neural decoder can be loaded into the processing unit for each iteration, which enables the hardware to be reused in different iterations.

\section{CONCLUSIONS}%问题、方法、不足、未来
In this correspondence, neural RC LDPC decoding networks that can deal with multiple code rates are proposed. The network exploits the structure of rate-compatible LDPC codes and dynamically activates partial neurons for different code rates during the decoding process. Experimental results show that neural decoding networks exhibit better FER performance than traditional decoding algorithms. The proposed neural RC decoder can replace multiple neural decoders without loss of FER performance.

%\section*{Acknowledgments}This should be a simple paragraph before the References to thank those individuals and institutions who have supported your work on this article.

%{\appendices
%\section*{Proof of the First Zonklar Equation}
%Appendix one text goes here.
% You can choose not to have a title for an appendix if you want by leaving the argument blank
%\section*{Proof of the Second Zonklar Equation}
%Appendix two text goes here.}

\bibliographystyle{IEEEtran}

\bibliography{reference.bib}

\vfill

\end{document}